\documentclass[aps,twocolumn,superscriptaddress]{revtex4}
\usepackage{epsfig}
\usepackage{times}

\begin{document}

\title{Collective behavior and evolutionary games -- An introduction}

\author{Matja{\v z} Perc}
\email{matjaz.perc@uni-mb.si}
\affiliation{Faculty of Natural Sciences and Mathematics, University of Maribor, Koro{\v s}ka cesta 160, SI-2000 Maribor, Slovenia}

\author{Paolo Grigolini}
\email{grigo@unt.edu}
\affiliation{Center for Nonlinear Science, University of North Texas, P.O. Box 311427, Denton, Texas 76203-1427, USA}

\maketitle

This is an introduction to the special issue titled ``Collective behavior and evolutionary games'' that is in the making at Chaos, Solitons \& Fractals. The term collective behavior covers many different phenomena in nature and society. From bird flocks and fish swarms to social movements and herding effects \cite{couzin_n07, castellano_rmp09, buchanan_np09, vicsek_pr12, ball_12}, it is the lack of a central planner that makes the spontaneous emergence of sometimes beautifully ordered and seemingly meticulously designed behavior all the more sensational and intriguing. The goal of the special issue is to attract submissions that identify unifying principles that describe the essential aspects of collective behavior, and which thus allow for a better interpretation and foster the understanding of the complexity arising in such systems. As the title of the special issue suggests, the later may come from the realm of evolutionary games, but this is certainly not a necessity, neither for this special issue, and certainly not in general. Interdisciplinary work on all aspects of collective behavior, regardless of background and motivation, and including synchronization \cite{pikovsky_03, acebron2005kuramoto, arenas_pr08} and human cognition \cite{baronchelli_tcs13}, is very welcome.

\section*{1. Evolutionary games}
Evolutionary games \cite{maynard_82, weibull_95, hofbauer_98, skyrms_04, nowak_06, sigmund_10} are, nevertheless, particularly likely to display some form of collective behavior, especially when played on structured populations \cite{szabo_pr07, roca_plr09}, and hence have been chosen to co-headline the special issue. Some background information and basic considerations follow.

Consider that players can choose either to cooperate or to defect. Mutual cooperation yields the reward $R$ to both players, mutual defection leads to punishment $P$ of both players, while the mixed choice gives the cooperator the sucker's payoff $S$ and the defector the temptation $T$. Typically $R=1$ and $P=0$ are considered fixed, while the remaining two payoffs can occupy $-1 \leq S \leq 1$ and $0 \leq T \leq 2$. If $T>R>P>S$ we have the prisoner's dilemma game, while $T>R>S>P$ yields the snowdrift game \cite{doebeli_el05}. Without much loss of generality, this parametrization is often further simplified for the prisoner's dilemma game, so that $T=b$ is the only free parameter while $R=1$ and $P=S=0$ are left constant. However, since the condition $P>S$ is no longer fulfilled, this version is usually referred to as the weak prisoner's dilemma game. For the snowdrift game one can, in a similar fashion, introduce $r \in [0,1]$ such that $T=1+r$ and $S=1-r$, where $r$ is the cost-to-benefit ratio and constitutes a diagonal in the snowdrift quadrant of the $T-S$ plane.

In the prisoner's dilemma game defectors dominate cooperators, so that in well-mixed populations natural selection always favors the former. In the snowdrift game \cite{santos_md_jtb12}, on the other hand, a coexistence of cooperators and defectors is possible even under well-mixed conditions, and spatial structure may even hinder the evolution of cooperation \cite{hauert_n04}. The prisoner's dilemma is in fact the most stringent cooperative dilemma, where for cooperation to arise a mechanism for the evolution of cooperation is needed \cite{nowak_s06}. This leads us to the year 1992, when Nowak and May \cite{nowak_n92b} observed the spontaneous formation of cooperative clusters on a square lattice, which enabled cooperators to survive in the presence of defectors, even in the realm of the prisoner's dilemma game. The mechanism is now most frequently referred to as network reciprocity or spatial reciprocity, and it became very popular in the wake of the progress made in network science and related interdisciplinary fields of research \cite{albert_rmp02, boccaletti_pr06, vespignani_np12, barabasi_np12}. The popularity was amplified further by the discovery that scale-free networks provide a unifying framework for the evolution of cooperation \cite{santos_prl05} -- a finding that subsequently motivated research on many different interaction networks \cite{szabo_pr07}, including such that coevolve as the game evolves \cite{zimmermann_pre04, pacheco_prl06, gross_jrsi08, perc_bs10, holme_pr12}.

The prisoner's dilemma and the snowdrift game are examples of pairwise interaction games. At each instance of the game, two players engage and receive payoffs based on their strategies. However, there are also games that are governed by group interactions, the most frequently studied of which is the public goods game \cite{santos_n08}. The basic setup with cooperators and defectors as the two competing strategies on a lattice can be described as follows \cite{szolnoki_pre09c}. Initially, $N = L^2$ players are arranged into overlapping groups of size $G$ such that everyone is surrounded by its $k=G-1$ neighbors and thus belongs to $g=G$ different groups, where $L$ is the linear system size and $k$ the degree (or coordination number) of the lattice. Cooperators contribute a fixed amount $a$, normally considered being equal to $1$ without loss of generality, to the common pool while defectors contribute nothing. Finally, the sum of all contributions in each group is multiplied by the synergy factor $r>1$ and the resulting public goods are distributed equally amongst all the group members. Despite obvious similarities with the prisoner's dilemma game (note that a public goods game in a group of size $G$ corresponds to $G-1$ pairwise prisoner's dilemma interactions), the outcomes of the two game types may differ significantly, especially in the details of collective behavior emerging on structured populations \cite{perc_jrsi13}.

\section*{2. Strategic complexity and more games}
\begin{figure*}
\begin{center}
\includegraphics[width=16cm]{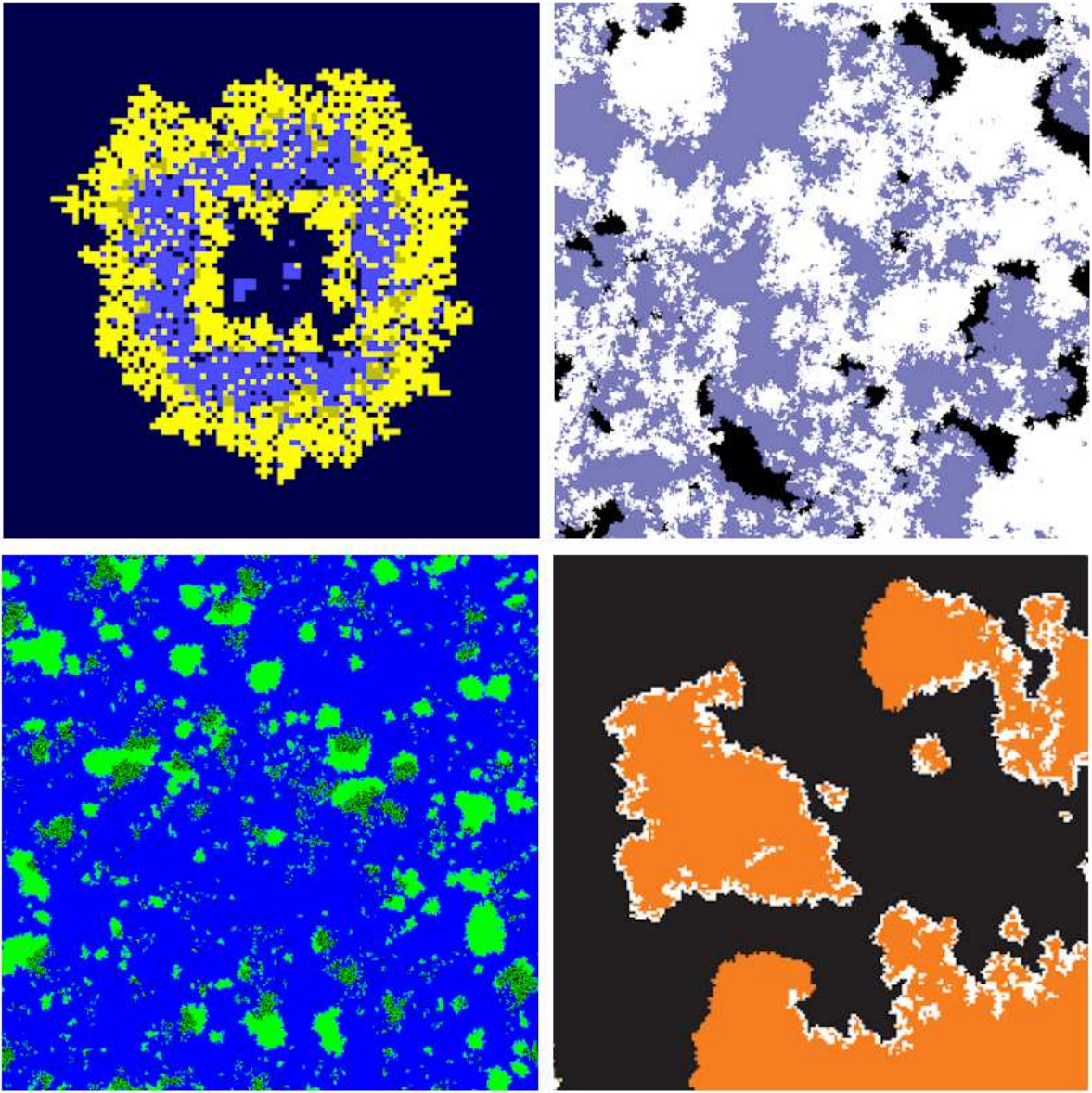}
\caption{Spatial patterns, emerging as a consequence of the spontaneous emergence of cyclic dominance between the competing strategies. Top left: Dynamically generated cyclic dominance in the spatial prisoner's dilemma game \cite{szolnoki_pre10b}. Light yellow (blue) are cooperators (defectors) whose learning capacity is minimal, while dark yellow (dark blue) are cooperators (defectors) whose learning capacity is maximal. Top right: Cyclic dominance in the spatial public goods game with pool-punishment \cite{szolnoki_pre11}. Black, white and blue are defectors ($D$), pure cooperator ($C$) and pool-punishers ($O$), respectively. Within the depicted (D+C+O)$_c$ phase, there are significantly different interfaces between the coexisting phases, which give rise to the anomalous ``survival of the weakest''. Pure cooperators behave as predators of pool-punishers, who in turn keep defectors in check, who in turn predate on pure cooperators. Bottom left: Cyclic dominance in the spatial ultimatum game with discrete strategies \cite{szolnoki_prl12}. The dominance is not between three strategies, but rather between two strategies ($E_1$ depicted blue and $E_2$ depicted green) and an alliance of two strategies ($E_{2}+A$, where $A$ is depicted black). Although similarly complex phases have been reported before in spatial ecological models \cite{szabo_jtb07} and in the spatial public goods game with pool punishment \cite{szolnoki_pre11}, the observation of qualitatively similar behavior in the ultimatum game enforces the notion that such exotic solutions may be significantly more common than initially assumed, especially in systems describing human behavior. Bottom right: Cyclical dominance between cooperators (white), defectors (black) and peer-punishers (orange) in the hard peer-punishment limit \cite{szolnoki_pre11b}. If punishment is sufficiently expensive and taxing on the defectors, this reduces the income of both defectors and peer-punishers. Along the interface, players can thus increase their payoff by choosing to cooperate, which manifests as the formation of white ``monolayers'' separating defectors and peer-punishers. We refer to the original works for further details about the studied evolutionary games.}
\label{cycles}
\end{center}
\end{figure*}

Significantly adding to the complexity of solutions are additional competing strategies that complement the traditional cooperators and defectors, such as loners or volunteers \cite{szabo_prl02, hauert_ajp05}, players that reward or punish \cite{sigmund_tee07, sigmund_pnas01, brandt_prsb03, rand_s09, helbing_ploscb10, szolnoki_epl10, sigmund_n10, szolnoki_pre11, rand_nc11, szolnoki_njp12, vukov_pcbi13}, or conditional cooperators and punishers \cite{szolnoki_jtb13, szolnoki_pre12}, to name but a few recently studied examples. These typically give rise to intricate phase diagrams, where continuous and discontinuous phase transitions delineate different stable solutions, ranging from single and two-strategy stationary states to rock-paper-scissors type cyclic dominance that can emerge in strikingly different ways. Figure~\ref{cycles} features characteristic snapshots of four representative examples.

Besides traditionally studied pairwise social dilemmas, such as the prisoner's dilemma and the snowdrift game, and the public goods game which is governed by group interactions, many other games have recently been studied as well. Examples include the related collective-risk social dilemmas \cite{santos_pnas11, chen_xj_epl12, moreira_jtb12, moreira_srep13} and stag-hunt dilemmas \cite{pacheco_prsb09}, as well as the ultimatum game \cite{guth_jebo82, nowak_s00, sigmund_sa02, kuperman_epjb08, eguiluz_acs09, sinatra_jstat09, xianyu_b_pa10b, gao_j_epl11, iranzo_jtb11, szolnoki_prl12}. Depending on the setup, most notably on whether the interactions among players are well-mixed or structured \cite{sigmund_n10, szolnoki_pre11}, but also on whether the strategy space is discrete or continuous \cite{nowak_s00, iranzo_jtb11, szolnoki_prl12}, these games exhibit equally complex behavior, and they invite further research along the lines outlined for the more traditionally studied evolutionary games described above.

\section*{3. Simulations versus reality}
Monte Carlo simulations are the predominant mode of analysis of evolutionary games on structured populations. Following the distribution of competing strategies uniformly at random, an elementary step entails randomly selecting a player and one of its neighbors, calculating the payoffs of both players, and finally attempting strategy adoption. The later is executed depending on the payoff difference, along with some uncertainty in the decision making to account for imperfect information and errors in judging the opponent. The temperature $K$ in the Fermi function \cite{szabo_pre98} is a popular choice to adjust the intensity of selection, and it is also frequently considered as a free parameter in determining the phase diagrams of games governed by pairwise interactions \cite{szabo_pre05} (note that for games governed by group interactions the impact of $K$ is qualitatively different and in fact less significant \cite{szolnoki_pre09c}). Repeating the elementary step $N$ times gives a chance once on average to every player to update its strategy, and thus constitutes one full Monte Carlo step.

Although simulations of games on structured populations are still far ahead of empirical studies and economic experiments \cite{camerer_03}, recent seminal advances based on large-scale human experiments suggest further efforts are needed to reconcile theory with reality \cite{grujic_pone12, gracia-lazaro_pnas12, gracia-lazaro_srep12}. According to the latter, network reciprocity does not account for why we so often choose socially responsible actions over defection, at least not in the realm of the prisoner's dilemma game. On the other hand, there is also evidence in support of cooperative behavior in human social networks \cite{fowler_pnas10, apicella_n12}, as well as in support of the fact that dynamic social networks do promote cooperation in experiments with humans \cite{rand_pnas11}. These findings, together with the massive amount of theoretical work that has been published in the past decades, promise exciting times ahead. Our hope is that this special issue will successfully capture some of this vibrancy and excitement, and in doing so hopefully recommend the journal to both readers and prospective authors.

\section*{4. Future research}
In terms of advisable future directions for research, at least in terms of evolutionary games, interdependent (or multiplex) networks certainly deserve mentioning. Not only are our social interactions limited and thus best described by models entailing networks rather than by well-mixed models, it is also a fact that these networks are often interdependent. It has recently been shown that even seemingly irrelevant changes in one network can have catastrophic and very much unexpected consequences in another network \cite{buldyrev_n10, gao_jx_np12, baxter_prl12, havlin_pst12, helbing_n13}, and since the evolution of cooperation in human societies also proceeds on such interdependent networks, it is of significant interest to determine to what extent the interdependence influences the outcome of evolutionary games. Existing works that have studied the evolution of cooperation on interdependent networks concluded that the interdependence can be exploited successfully to promote cooperation \cite{wang_z_epl12, gomez-gardenes_srep12, gomez-gardenes_pre12, wang_b_jsm12}, for example through the means of interdependent network reciprocity \cite{wang_z_srep13} or information sharing \cite{szolnoki_njp13}, but also that too much interdependence is not good either. In particular, individual networks must also be sufficiently independent to remain functional if the evolution of cooperation in the other network goes terribly wrong. Also of interest are evolutionary games on bipartite networks \cite{gomez-gardenes_c11, gomez-gardenes_epl11}, where group structure is considered separately from the network structure, and thus enables a deeper understanding of the evolution of cooperation in games that are governed by group interactions. It seems that the evolution of cooperation on both interdependent and bipartite networks has reached fruition to a degree that the next step might be to consider coevolution between cooperation and either interdependence or bipartiteness. Lastly we also refer to \cite{nowak_jtb12}, where Section 4 features 10 interesting open problems that certainly merit attention.

Finally, we would like the potential authors to explore also the challenging issue of a possible connection between sociology and neurophysiology.  In same cases \cite{szabo_prl02} game theory, which is widely applied in sociology,  generates patterns reminiscent of those produced by the Ising model thereby suggesting a possible connection with criticality \cite{originalchialvo}, which is becoming an increasingly popular hypothesis in neurophysiology, especially for brain dynamics \cite{chialvo,plenz}, where this assumption generates theoretical results yielding  a surprisingly good agreement with the experimental observation of real brain \cite{chialvoprl}. The potential authors may also contribute significant advances to understand the real nature of  neurophysiological criticality, whose connections with the criticality of physical systems are not yet satisfactorily established \cite{elisa}, although criticality-induced dynamics are proven to be responsible for a network evolution fitting the main subject of this special issue as well as the crucial neurophysiological hypothesis of Hebbian learning \cite{gosia1}. The decision making model \cite{gosia1}, a dynamical model sharing \cite{gosia2} the same criticality properties as those adopted to study the brain dynamics \cite{originalchialvo}, generates an interesting phenomenon  that the authors in \cite{gosia3} used to explain the Arab Spring events. This is a sociological phenomenon, where a small number of individual produces substantial changes in social consensus \cite{gosia3}, in agreement with similar results based on the adoption of game theory \cite{possibleconnection}. The theoretical reasons of this surprising agreement is one of the problems that hopefully some contributors to this special issue may solve. We hope that papers on this issue may help to establish a connection between criticality \cite{vanni} and swarm intelligence \cite{couzin_n07} and hopefully between cognition \cite{baronchelli_tcs13} and consciousness \cite{allegro}.

To conclude, we note that this special issue is also about to feature future research. In order to avoid delays that are sometimes associated with waiting for a special issue to become complete before it is published, we have adopted an alternative approach. The special issue will be updated continuously from the publication of this introduction onwards, meaning that new papers will be published immediately after acceptance. The issue will hopefully grow in size on a regular basis, with the last papers being accepted no later than August 30th for the special issue to be closed by the end of 2013. The down side of this approach is that we cannot feature the traditional brief summaries of each individual work that will be published, but we hope that this is more than made up for by the immediate availability of the latest research. Please stay tuned, and consider contributing to ``Collective behavior and evolutionary games''.

\begin{acknowledgments}
We would like to thank the Editors-in-Chief of Chaos, Solitons \& Fractals, Stefano Boccaletti and Maurice Courbage, to support this special issue dedicated to the collective behavior and evolutionary games. We would also like to thank all the authors and referees for their valuable contributions and help.
\end{acknowledgments}


\begin{thebibliography}{10}
\expandafter\ifx\csname url\endcsname\relax
  \def\url#1{\texttt{#1}}\fi
\expandafter\ifx\csname urlprefix\endcsname\relax\def\urlprefix{URL }\fi
\expandafter\ifx\csname href\endcsname\relax
  \def\href#1#2{#2} \def\path#1{#1}\fi

\bibitem{couzin_n07}
I.~Couzin, Collective minds, Nature 445 (2007) 715--715.

\bibitem{castellano_rmp09}
C.~Castellano, S.~Fortunato, V.~Loreto, Statistical physics of social dynamics,
  Rev. Mod. Phys. 81 (2009) 591--646.

\bibitem{buchanan_np09}
M.~Buchanan, Collectivist revolution in evolution, Nat. Phys. 5 (2009)
  531--531.

\bibitem{vicsek_pr12}
T.~Vicsek, A.~Zafeiris, Collective motion, Phys. Rep. 517 (2012) 71--140.

\bibitem{ball_12}
P.~Ball, D.~Helbing, Why Society is a Complex Matter, Springer, Berlin
  Heidelberg, 2012.

\bibitem{pikovsky_03}
A.~Pikovsky, M.~Rosenblum, J.~Kurths, Synchronization: a universal concept in
  nonlinear sciences, Cambridge University Press, Cambridge, 2003.

\bibitem{acebron2005kuramoto}
J.~A. Acebr{\'o}n, L.~L. Bonilla, C.~J. P{\'e}rez Vicente, F. Ritort, R. Spigler, The Kuramoto model: A simple paradigm for synchronization phenomena, Rev. Mod. Phys. 77 (2005) 137--185.

\bibitem{arenas_pr08}
A.~Arenas, A.~D{\'i}az-Guilera, J.~Kurths, Y.~Moreno, C.~Zhou, Synchronization
  in complex networks, Phys. Rep. 469 (2008) 93--153.

\bibitem{baronchelli_tcs13}
A.~Baronchelli, R.~\protect{Ferrer-i-Cancho}, R.~\protect{Pastor-Satorras},
  N.~Chater, M.~H. Christiansen, Networks in cognitive science, Trends in
  Cognitive Sciences (2013) in press.

\bibitem{maynard_82}
J.~Maynard~Smith, Evolution and the Theory of Games, Cambridge University
  Press, Cambridge, U.K., 1982.

\bibitem{weibull_95}
J.~W. Weibull, Evolutionary Game Theory, MIT Press, Cambridge, MA, 1995.

\bibitem{hofbauer_98}
J.~Hofbauer, K.~Sigmund, Evolutionary Games and Population Dynamics, Cambridge
  University Press, Cambridge, 1998.

\bibitem{skyrms_04}
B.~Skyrms, Stag-Hunt Game and the Evolution of Social Structure, Cambridge
  University Press, Cambridge, 2004.

\bibitem{nowak_06}
M.~A. Nowak, Evolutionary Dynamics, Harvard University Press, Cambridge, MA,
  2006.

\bibitem{sigmund_10}
K.~Sigmund, The Calculus of Selfishness, Princeton University Press, Princeton,
  NJ, 2010.

\bibitem{szabo_pr07}
G.~Szab{\'o}, G.~F{\'a}th, Evolutionary games on graphs, Phys. Rep. 446 (2007)
  97--216.

\bibitem{roca_plr09}
C.~P. Roca, J.~A. Cuesta, A.~S{\'a}nchez, Evolutionary game theory: Temporal
  and spatial effects beyond replicator dynamics, Phys. Life Rev. 6 (2009)
  208--249.

\bibitem{doebeli_el05}
M.~Doebeli, C.~Hauert, Models of cooperation based on prisoner's dilemma and
  snowdrift game, Ecol. Lett. 8 (2005) 748--766.

\bibitem{santos_md_jtb12}
M.~Santos, F.~Pinheiro, F.~Santos, J.~Pacheco, Dynamics of $n$-person snowdrift
  games in structured populations, J. Theor. Biol. 315 (2012) 81--86.

\bibitem{hauert_n04}
C.~Hauert, M.~Doebeli, Spatial structure often inhibits the evolution of
  cooperation in the snowdrift game, Nature 428 (2004) 643--646.

\bibitem{nowak_s06}
M.~A. Nowak, Five rules for the evolution of cooperation, Science 314 (2006)
  1560--1563.

\bibitem{nowak_n92b}
M.~A. Nowak, R.~M. May, Evolutionary games and spatial chaos, Nature 359 (1992)
  826--829.

\bibitem{albert_rmp02}
R.~Albert, A.-L. Barab{\'a}si, Statistical mechanics of complex networks, Rev.
  Mod. Phys. 74 (2002) 47--97.

\bibitem{boccaletti_pr06}
S.~Boccaletti, V.~Latora, Y.~Moreno, M.~Chavez, D.~Hwang, Complex networks:
  Structure and dynamics, Phys. Rep. 424 (2006) 175--308.

\bibitem{vespignani_np12}
A.~Vespignani, Modelling dynamical processes in complex socio-technical
  systems, Nat. Phys. 8 (2012) 32--–39.

\bibitem{barabasi_np12}
A.~L. Barab{\'a}si, The network takeover, Nat. Phys. 8 (2012) 14--16.

\bibitem{santos_prl05}
F.~C. Santos, J.~M. Pacheco, Scale-free networks provide a unifying framework
  for the emergence of cooperation, Phys. Rev. Lett. 95 (2005) 098104.

\bibitem{zimmermann_pre04}
M.~G. Zimmermann, V.~Egu{\'{\i}}luz, M.~S. Miguel, Coevolution of dynamical
  states and interactions in dynamic networks, Phys. Rev. E 69 (2004)
  065102(R).

\bibitem{pacheco_prl06}
J.~M. Pacheco, A.~Traulsen, M.~A. Nowak, Coevolution of strategy and structure
  in complex networks with dynamical linking, Phys. Rev. Lett. 97 (2006)
  258103.

\bibitem{gross_jrsi08}
T.~Gross, B.~Blasius, Adaptive coevolutionary networks: a review, J. R. Soc.
  Interface 5 (2008) 259--271.

\bibitem{perc_bs10}
M.~Perc, A.~Szolnoki, Coevolutionary games -- a mini review, BioSystems 99
  (2010) 109--125.

\bibitem{holme_pr12}
P.~Holme, J.~Saram{\"a}ki, Temporal networks, Phys. Rep. 519 (2012) 97--125.

\bibitem{santos_n08}
F.~C. Santos, M.~D. Santos, J.~M. Pacheco, Social diversity promotes the
  emergence of cooperation in public goods games, Nature 454 (2008) 213--216.

\bibitem{szolnoki_pre09c}
A.~Szolnoki, M.~Perc, G.~Szab{\'o}, Topology-independent impact of noise on
  cooperation in spatial public goods games, Phys. Rev. E 80 (2009) 056109.

\bibitem{perc_jrsi13}
M.~Perc, J.~G{\'o}mez-Garde{\~n}es, A.~Szolnoki, L.~M. Flor{\'{\i}a and Y.
  Moreno}, Evolutionary dynamics of group interactions on structured
  populations: a review, J. R. Soc. Interface 10 (2013) 20120997.

\bibitem{szabo_prl02}
G.~Szab{\'o}, C.~Hauert, Phase transitions and volunteering in spatial public
  goods games, Phys. Rev. Lett. 89 (2002) 118101.

\bibitem{hauert_ajp05}
C.~Hauert, G.~Szab{\'o}, Game theory and physics, Am. J. Phys. 73 (2005)
  405--414.

\bibitem{sigmund_tee07}
K.~Sigmund, Punish or perish? retailation and collaboration among humans,
  Trends Ecol. Evol. 22 (2007) 593--600.

\bibitem{sigmund_pnas01}
K.~Sigmund, C.~Hauert, M.~A. Nowak, Reward and punishment, Proc. Natl. Acad.
  Sci. USA 98 (2001) 10757--10762.

\bibitem{brandt_prsb03}
H.~Brandt, C.~Hauert, K.~Sigmund, Punishment and reputation in spatial public
  goods games, Proc. R. Soc. Lond. B 270 (2003) 1099--1104.

\bibitem{rand_s09}
D.~G. Rand, A.~Dreber, T.~Ellingsen, D.~Fudenberg, M.~A. Nowak, Positive
  interactions promote public cooperation, Science 325 (2009) 1272--1275.

\bibitem{helbing_ploscb10}
D.~Helbing, A.~Szolnoki, M.~Perc, G.~Szab{\'o}, Evolutionary establishment of
  moral and double moral standards through spatial interactions, PLoS Comput.
  Biol. 6 (2010) e1000758.

\bibitem{szolnoki_epl10}
A.~Szolnoki, M.~Perc, Reward and cooperation in the spatial public goods game,
  EPL 92 (2010) 38003.

\bibitem{sigmund_n10}
K.~Sigmund, H.~De~Silva, A.~Traulsen, C.~Hauert, Social learning promotes
  institutions for governing the commons, Nature 466 (2010) 861--863.

\bibitem{szolnoki_pre11}
A.~Szolnoki, G.~Szab{\'o}, M.~Perc, Phase diagrams for the spatial public goods
  game with pool punishment, Phys. Rev. E 83 (2011) 036101.

\bibitem{rand_nc11}
D.~G. Rand, M.~A. Nowak, The evolution of antisocial punishment in optional
  public goods games, Nat. Commun. 2 (2011) 434.

\bibitem{szolnoki_njp12}
A.~Szolnoki, M.~Perc, Evolutionary advantages of adaptive rewarding, New J.
  Phys. 14 (2012) 093016.

\bibitem{vukov_pcbi13}
J.~Vukov, F.~Pinheiro, F.~Santos, J.~Pacheco, Reward from punishment does not
  emerge at all costs, PLoS Comput. Biol. 9 (2013) e1002868.

\bibitem{szolnoki_jtb13}
A.~Szolnoki, M.~Perc, Effectiveness of conditional punishment for the evolution
  of public cooperation, J. Theor. Biol. 325 (2013) 34--41.

\bibitem{szolnoki_pre12}
A.~Szolnoki, M.~Perc, Conditional strategies and the evolution of cooperation
  in spatial public goods games, Phys. Rev. E 85 (2012) 026104.

\bibitem{santos_pnas11}
F.~C. Santos, J.~M. Pacheco, Risk of collective failure provides an escape from
  the tragedy of the commons, Proc. Natl. Acad. Sci. USA 108 (2011)
  10421--10425.

\bibitem{chen_xj_epl12}
X.~Chen, A.~Szolnoki, M.~Perc, Averting group failures in collective-risk
  social dilemmas, EPL 99 (2012) 68003.

\bibitem{moreira_jtb12}
J.~Moreira, F.~Pinheiro, A.~Nunes, J.~Pacheco, Evolutionary dynamics of
  collective action when individual fitness derives from group decisions taken
  in the past, J. Theor. Biol. 298 (2012) 8--15.

\bibitem{moreira_srep13}
J.~Moreira, J.~M. Pacheco, F.~C. Santos, Evolution of collective action in
  adaptive social structures, Sci. Rep. 3 (2013) 1521.

\bibitem{pacheco_prsb09}
J.~M. Pacheco, F.~C. Santos, M.~O. Souza, B.~Skyrms, Evolutionary dynamics of
  collective action in $n$-person stag hunt dilemmas, Proc. R. Soc. Lond. B 276
  (2009) 315--321.

\bibitem{guth_jebo82}
W.~G{\"u}th, R.~Schmittberger, B.~Schwarze, An experimental analysis of
  ultimatum bargaining, J. Econ. Behav. Org. 3 (1982) 367--388.

\bibitem{nowak_s00}
M.~A. Nowak, K.~M. Page, K.~Sigmund, Fairness versus reason in the ultimatum
  game, Science 289 (2000) 1773--1775.

\bibitem{sigmund_sa02}
K.~Sigmund, E.~Fehr, M.~A. Nowak, The economics of fair play, Sci. Am. 286
  (2002) 82--87.

\bibitem{kuperman_epjb08}
M.~N. Kuperman, S.~Risau-Gusman, The effect of topology on the spatial
  ultimatum game, Eur. Phys. J. B 62 (2008) 233--238.

\bibitem{eguiluz_acs09}
V.~M. Equ{\'{\i}}luz, C.~Tessone, Critical behavior in an evolutionary
  ultimatum game with social structure, Adv. Complex Systems 12 (2009)
  221--232.

\bibitem{sinatra_jstat09}
R.~Sinatra, J.~Iranzo, J.~G{\'o}mez-Garde{\~n}es, L.~M. Flor\'{\i}a, V.~Latora,
  Y.~Moreno, The ultimatum game in complex networks, J. Stat. Mech. (2009)
  P09012.

\bibitem{xianyu_b_pa10b}
B.~Xianyu, J.~Yang, Evolutionary ultimatum game on complex networks under
  incomplete information, Physica A 389 (2010) 1115--1123.

\bibitem{gao_j_epl11}
J.~Gao, Z.~Li, T.~Wu, L.~Wang, The coevolutionary ultimatum game, EPL 93 (2011)
  48003.

\bibitem{iranzo_jtb11}
J.~Iranzo, J.~Rom{\'a}n, A.~S{\'a}nchez, The spatial ultimatum game revisited,
  J. Theor. Biol. 278 (2011) 1--10.

\bibitem{szolnoki_prl12}
A.~Szolnoki, M.~Perc, G.~Szab{\'o}, Defense mechanisms of empathetic players in
  the spatial ultimatum game, Phys. Rev. Lett. 109 (2012) 078701.

\bibitem{szabo_pre98}
G.~Szab{\'o}, C.~T{\H{o}}ke, Evolutionary prisoner's dilemma game on a square
  lattice, Phys. Rev. E 58 (1998) 69--73.

\bibitem{szabo_pre05}
G.~Szab{\'o}, J.~Vukov, A.~Szolnoki, Phase diagrams for an evolutionary
  prisoner's dilemma game on two-dimensional lattices, Phys. Rev. E 72 (2005)
  047107.

\bibitem{camerer_03}
C.~F. Camerer, Behavioral Game Theory: Experiments in Strategic Interaction,
  Princeton University Press, Princeton, 2003.

\bibitem{grujic_pone12}
J.~Gruji{\'c}, T.~R{\"o}hl, D.~Semmann, M.~Milinksi, A.~Traulsen, Consistent
  strategy updating in spatial and non-spatial behavioral experiments does not
  promote cooperation in social networks, PLoS ONE 7 (2012) e47718.

\bibitem{gracia-lazaro_pnas12}
C.~Gracia-L{\'a}zaro, A.~Ferrer, G.~Ruiz, A.~Taranc{\'o}n, J.~Cuesta,
  A.~S{\'a}nchez, Y.~Moreno, Heterogeneous networks do not promote cooperation
  when humans play a prisoner's dilemma, Proc. Natl. Acad. Sci. USA 109 (2012)
  12922--12926.

\bibitem{gracia-lazaro_srep12}
C.~Gracia-L{\'a}zaro, J.~Cuesta, A.~S{\'a}nchez, Y.~Moreno, Human behavior in
  prisoner's dilemma experiments suppresses network reciprocity, Sci. Rep. 2
  (2012) 325.

\bibitem{fowler_pnas10}
J.~H. Fowler, N.~A. Christakis, Cooperative behavior cascades in human social
  networks, Proc. Natl. Acad. Sci. USA 107 (2010) 5334--5338.

\bibitem{apicella_n12}
C.~L. Apicella, F.~W. Marlowe, J.~H. Fowler, N.~A. Christakis, Social networks
  and cooperation in hunter-gatherers, Nature 481 (2012) 497--501.

\bibitem{rand_pnas11}
D.~G. Rand, S.~Arbesman, N.~A. Christakis, Dynamic social networks promote
  cooperation in experiments with humans, Proc. Natl. Acad. Sci. USA 108 (2011)
  19193--19198.

\bibitem{szolnoki_pre10b}
A.~Szolnoki, Z.~Wang, J.~Wang, X.~Zhu, Dynamically generated cyclic dominance
  in spatial prisoner's dilemma games, Phys. Rev. E 82 (2010) 036110.

\bibitem{szabo_jtb07}
P.~Szab{\'o}, T.~Cz{\'a}r{\'a}n, G.~Szab{\'o}, Competing associations in
  bacterial warfare with two toxins, J. Theor. Biol. 248 (2007) 736--744.

\bibitem{szolnoki_pre11b}
A.~Szolnoki, G.~Szab{\'o}, L.~Czak{\'o}, Competition of individual and
  institutional punishments in spatial public goods games, Phys. Rev. E 84
  (2011) 046106.

\bibitem{buldyrev_n10}
S.~V. Buldyrev, R.~Parshani, G.~Paul, H.~E. Stanley, S.~Havlin, Catastrophic
  cascade of failures in interdependent networks, Nature 464 (2010) 1025--1028.

\bibitem{gao_jx_np12}
J.~Gao, S.~V. Buldyrev, H.~E. Stanley, S.~Havlin, Networks formed from
  interdependent networks, Nat. Phys. 8 (2012) 40--48.

\bibitem{baxter_prl12}
G.~J. Baxter, S.~N. Dorogovtsev, A.~V. Goltsev, J.~F.~F. Mendes, Avalanche
  collapse of interdependent networks, Phys. Rev. Lett. 109 (2012) 248701.

\bibitem{havlin_pst12}
S.~Havlin, D.~Y. Kenett, E.~\protect{Ben-Jacob}, A.~Bunde, H.~Hermann,
  J.~Kurths, S.~Kirkpatrick, S.~Solomon, J.~Portugali, Challenges of network
  science: Applications to infrastructures, climate, social systems and
  economics, Eur. J. Phys. Special Topics 214 (2012) 273--293.

\bibitem{helbing_n13}
D.~Helbing, Globally networked risks and how to respond, Nature 497 (2013)
  51--59.

\bibitem{wang_z_epl12}
Z.~Wang, A.~Szolnoki, M.~Perc, Evolution of public cooperation on
  interdependent networks: The impact of biased utility functions, EPL 97
  (2012) 48001.

\bibitem{gomez-gardenes_srep12}
J.~G{\'o}mez-Garde{\~n}es, I.~Reinares, A.~Arenas, L.~M. Flor{\' \i}a,
  Evolution of cooperation in multiplex networks, Sci. Rep. 2 (2012) 620.

\bibitem{gomez-gardenes_pre12}
J.~G{\'o}mez-Garde{\~n}es, C.~Gracia-L{\'a}zaro, L.~M. Flor{\' \i}a, Y.~Moreno,
  Evolutionary dynamics on interdependent populations, Phys. Rev. E 86 (2012)
  056113.

\bibitem{wang_b_jsm12}
B.~Wang, X.~Chen, L.~Wang, Probabilistic interconnection between interdependent
  networks promotes cooperation in the public goods game, J. Stat. Mech. (2012) P11017.

\bibitem{wang_z_srep13}
Z.~Wang, A.~Szolnoki, M.~Perc, Interdependent network reciprocity in
  evolutionary games, Sci. Rep. 3 (2013) 1183.

\bibitem{szolnoki_njp13}
A.~Szolnoki, M.~Perc, Information sharing promotes prosocial behaviour, New J.
  Phys. 15 (2013) 053010.

\bibitem{gomez-gardenes_c11}
J.~G{\'o}mez-Garde{\~n}es, M.~Romance, R.~Criado, D.~Vilone, A.~S{\'a}nchez,
  Evolutionary games defined at the network mesoscale: The public goods game,
  Chaos 21 (2011) 016113.

\bibitem{gomez-gardenes_epl11}
J.~G{\'o}mez-Garde{\~n}es, D.~Vilone, A.~S{\'a}nchez, Disentangling social and
  group heterogeneities: Public goods games on complex networks, EPL 95 (2011)
  68003.

\bibitem{nowak_jtb12}
M.~A. Nowak, Evolving cooperation, J. Theor. Biol. 299 (2012) 1--8.

\bibitem{originalchialvo}
D. Fraiman, P. Balenzuela J. Foss, D.~R. Chialvo, Ising-like dynamics in large-scale functional brain networks, Phys. Rev. E 79 (2009) 061922.

\bibitem{chialvo}
P. Grigolini and D.~R. Chialvo (Guest Editors), Chaos, Solitons \& Fractals Special Issue: Emergent Critical Brain Dynamics, 2013.

\bibitem{plenz}
D. Plenz (Editor), Criticality in the Brain, John Wiley \& Sons, 2013.

\bibitem{chialvoprl}
A. Haimovici, E. Tagliazucchi, P. Balenzuela, D.~R. Chialvo, Phys. Rev. Lett. 110 (2013) 178101.

\bibitem{elisa}
E. Lovecchio, P. Allegrini, E. Geneston, B.~J. West, P. Grigolini, From self-organized to extended criticality, Front. Physiol. 3 (2012) 98.

\bibitem{gosia1}
M. Turalska, E. Geneston, B.~J. West, P. Allegrini, P. Grigolini, Cooperation-Induced Topological Complexity: A Promising Road to Fault Tolerance and Hebbian Learning, Front. Physiol. 3 (2012)  52.

\bibitem{gosia2}
B.~J. West, M. Turalska, P. Grigolini, From Social Crises to Neuronal Avalanches, in Criticality in the Brain, D. Plenz (Editor), John Wiley \& Sons, 2013.

\bibitem{gosia3}
M. Turalska, B.~J. West, P. Grigolini, Role of committed minorities in times of crisis, Sci. Rep. 3 (2013) 1371.

\bibitem{possibleconnection}
P. Sing, S. Sreenivasan, B.~K. Szymanski, G. Korniss, Accelerating consensus on coevolving networks: The effect of committed individuals, Phys. Rev. E 85 (2012) 046104.

\bibitem{vanni}
F. Vanni, M. Lukovic, P. Grigolini, Criticality and Transmission of Information in a Swarm of Cooperative Units, Phys. Rev. Lett. 107 (2012) 041145.

\bibitem{allegro}
P.  Allegrini, P. Paradisi, D. Menicucci, M. Laurino, R. Bedini, A. Piarulli, A. Gemignani, Sleep unconsciousness and breakdown of serial critical intermittency: New vistas on the global workspace, in Chaos, Solitons \& Fractals Special Issue: Emergent Critical Brain Dynamics, P. Grigolini and D. R. Chialvo (Guest Editors), 2013.

\end{thebibliography}
\end{document}